\documentclass[10pt]{emulateapj}
\usepackage{amsmath}
\usepackage{bm}
\usepackage{graphicx}
\usepackage{natbib}
\begin{document}

\title{Intermittency and Alignment in Strong RMHD Turbulence}

\author{B. D. G. Chandran\altaffilmark{1,2}, 
 A. A. Schekochihin\altaffilmark{2,3}, and A. Mallet\altaffilmark{3}}

\altaffiltext{1}{Space Science Center and Department of Physics,  University of New Hampshire, Durham, NH 03824;  benjamin.chandran@unh.edu}

\altaffiltext{2}{Merton College, University of Oxford, Oxford OX1 4JD, United Kingdom}

\altaffiltext{3}{Rudolf Peierls Centre for Theoretical Physics, University of Oxford, Oxford OX1 3NP, United Kingdom}

\begin{abstract}
We develop an analytic model of intermittent, three-dimensional,
strong, reduced magnetohydrodynamic (RMHD) turbulence with zero cross
helicity. We take the fluctuation amplitudes to have a log-Poisson
distribution and incorporate into the model a new phenomenology of
scale-dependent dynamic alignment between the Els\"asser
variables~$\displaystyle \bm{z}^\pm $.  We find that the structure
function $\displaystyle \langle |\Delta \bm{z}^\pm _\lambda|^n\rangle$
scales as $\displaystyle \lambda^{1-\beta^n}$, where $\displaystyle
\Delta \bm{z}^\pm_\lambda$ is the variation in~$\displaystyle
\bm{z}^\pm$ across a distance~$\displaystyle \lambda$ perpendicular to
the magnetic field. We calculate the value of~$\beta$ to be~$\simeq
0.69$ based on our assumptions that the energy cascade rate is
independent of~$\displaystyle \lambda$ within the inertial range, that
the most intense coherent structures are two-dimensional with a volume
filling factor~$\propto \lambda$, and that most of the cascade power
arises from interactions between exceptionally intense fluctuations
and much weaker fluctuations.  Two consequences of this
structure-function scaling are that the total-energy power spectrum is
$\displaystyle \propto k_\perp^{-1.52}$ and that the kurtosis of the
fluctuations is~$\displaystyle \propto\lambda^{-0.27}$.  Our model
resolves the problem that alignment angles defined in different ways
exhibit different scalings. Specifically, we find that the
energy-weighted average angle between the velocity and magnetic-field
fluctuations is $\displaystyle \propto \lambda^{0.21}$, the
energy-weighted average angle between~$\displaystyle \Delta \bm{z}^+$
and $\displaystyle \Delta \bm{z}^-$ is~$\displaystyle \propto
\lambda^{0.10}$, and the average angle between~$\displaystyle \Delta
\bm{z}^+$ and $\displaystyle \Delta \bm{z}^-$ without energy weighting
is~$\propto [\ln(L/\lambda)]^{-1/2}$ when $L/\lambda \gg 1$, where $L$
is the outer scale.  These scalings appear to be consistent with
numerous results from direct numerical simulations.
\end{abstract} 

\maketitle

\vspace{0.0cm} 
\section{Introduction}
\label{sec:intro}
\vspace{0.0cm} 

Plasma turbulence plays an important role in many astrophysical
systems, including accretion flows around black holes, intracluster
plasmas in clusters of galaxies, and outflows from stars, including
the solar wind.  In many of these systems, the energetically dominant
component of the turbulence is non-compressive and can be modeled, at
least in an approximate way, within the framework of incompressible
magnetohydrodynamics (MHD).

In incompressible MHD, velocity and magnetic-field fluctuations
($\displaystyle \delta \bm{v}$ and $\displaystyle \delta \bm{B}$)
propagate either parallel or anti-parallel to the local background
magnetic field~$\displaystyle \bm{B}_{\rm loc}$, and nonlinear
interactions occur only between counter-propagating
fluctuations~\citep{iroshnikov63,kraichnan65}. As a consequence, the
energy cascade is anisotropic, producing small-scale structures or
``eddies'' that satisfy $\lambda \ll l$, where $l$ ($\lambda$) is the
correlation length of an eddy parallel (perpendicular) to~$\bm{B}_{\rm
  loc}$~\citep{shebalin83,goldreich95,ng96,galtier00,cho00,maron01}.
When $\lambda \ll l$, the components of $\delta \bm{v}$ and $\delta
\bm{B}$ perpendicular to~$\bm{B}_{\rm loc}$ evolve independently of
the components parallel to~$\bm{B}_{\rm loc}$ and are well described
by reduced MHD (RMHD)~\citep{kadomtsev74,strauss76}. When $\delta B
\ll B_{\rm loc}$ and $\rho_{\rm p} \ll \lambda \ll l$, where
$\rho_{\rm p}$ is the proton gyroradius, RMHD is a rigorous limit of
gyrokinetics and is valid for both collisional and collisionless
plasmas~\citep{schekochihin09}.

In this paper, we propose a phenomenological theory of RMHD turbulence
that goes beyond scaling theories for
spectra~\citep{iroshnikov63,kraichnan65,goldreich95,boldyrev06} and
allows us to make predictions concerning the scale dependence of
arbitrary-order structure functions and the relative orientation of
the turbulent magnetic field and velocity. A new feature of this
theory is that it accounts, within one model, for both intermittency
and scale-dependent dynamic alignment (SDDA).

The concept of SDDA was introduced by \cite{boldyrev05,boldyrev06},
who argued that the angle $\phi_\lambda$ between~$\delta
\bm{v}_\lambda$ and $\delta \bm{B}_\lambda$ decreases with
decreasing~$\lambda$, where $\delta \bm{v}_{\lambda}$ and $\delta
\bm{B}_\lambda$ are the fluctuations in the velocity and magnetic
field at perpendicular scale~$\lambda$.  As $\phi_\lambda$ decreases,
nonlinear interactions in RMHD weaken, causing the power spectrum of
the fluctuation energy to flatten relative to models that
neglect~SDDA.

Intermittency is the phenomenon in which the fluctuation energy is
concentrated into an increasingly small fraction of the volume
as~$\lambda \rightarrow 0$.  Intermittency has been measured in
hydrodynamic turbulence~\citep[e.g.,][]{benzi93}, solar-wind
turbulence~\citep{burlaga91,horbury97,sorriso99,forman03,bruno07,wan12b,osman12,perri12,osman14},
numerical simulations of MHD turbulence and RMHD
turbulence~\citep{muller00,maron01,muller03,beresnyak06,mininni09,imazio13},
and hybrid-Vlasov and particle-in-cell simulations of plasma
turbulence~\citep{greco12,servidio12,wan12,karimabadi13,wu13}.  A
number of theoretical models have been introduced to describe
intermittency, including the log-normal
model~\citep{kolmogorov62,gurvich67}, the ``constant-$\beta$''
model~\citep{frisch78}, and multi-fractal models in which the
fluctuation amplitudes scale differently on different subsets of the
volume that have different fractal
dimensions~\citep{parisi85,paladin87}.  One such multi-fractal model,
based on a log-Poisson probability distribution function for the local
dissipation rate, was developed by \cite{she94} \citep[see
  also][]{dubrulle94}.  She \& Leveque's~(1994) approach served as the
basis for several previous studies of intermittency in both
compressible and incompressible MHD
turbulence~\citep{grauer94,politano95,muller00,boldyrev02a,boldyrev02b}.

We draw upon ideas from the She-Leveque model to construct an analytic
model of strong RMHD turbulence that incorporates a new phenomenology
of SDDA. We present this model in Section~\ref{sec:theory}.  In
Section~\ref{sec:comp}, we compare our model with previously published
numerical simulations, and in Section~\ref{sec:discussion} we discuss
our results and the relation between our work and previous turbulence
models.

\vspace{0.0cm} 
\section{Analytic Model of Strong RMHD Turbulence}
\label{sec:theory} 
\vspace{0.0cm}

The equations of incompressible MHD can be written in the form
\begin{equation}
\frac{\partial \bm{z}^\pm}{\partial t} \mp \bm{v}_{\rm A} \cdot \nabla
\bm{z}^\pm = - \bm{z}^\mp \cdot \nabla \bm{z}^\pm - \nabla \Pi,
\label{eq:RMHD} 
\end{equation} 
where $\bm{z}^\pm = \delta \bm{v} \pm \delta \bm{B}/\sqrt{4\pi \rho}$
are the Els\"asser variables, $\delta \bm{v}$ and $\delta \bm{B}$ are
the velocity and magnetic-field fluctuations, $\rho$ is the mass
density, $\bm{v}_{\rm A} = \bm{B}_0/\sqrt{4\pi \rho}$ is the Alfv\'en
velocity, $\bm{B}_0$ is the background magnetic field, $\Pi = (p +
B^2/8\pi)/\rho$, $p$ is the pressure, $\bm{B} = \bm{B}_0 + \delta
\bm{B}$, and $\nabla \cdot \bm{z}^\pm = 0$. The RMHD equations
are equivalent to Equation~(\ref{eq:RMHD}) supplemented by the condition
\begin{equation}
\bm{B}_0 \cdot \bm{z}^\pm = 0.
\label{eq:RMHDsuppl} 
\end{equation} 
Throughout this paper, we neglect dissipation and focus on the inertial
range.

\vspace{0.0cm} 
\subsection{Statistical Distribution of Field Increments}
\label{sec:statistical} 
\vspace{0.0cm}

We consider the turbulence to be an ensemble of approximately
localized $z^+$~and $z^-$~structures. We define 
\begin{equation}
\Delta \bm{z}^\pm_\lambda = \bm{z}^\pm(\bm{x} + 0.5\lambda
\bm{\hat{s}},t) - \bm{z}^\pm(\bm{x}-0.5\lambda\bm{\hat{s}},t)
\label{eq:Deltaz},
\end{equation}
where $\bm{\hat{s}}$ is a unit vector perpendicular
to~$\bm{B}(\bm{x},t)$.  We define $\delta z^\pm_\lambda$ to be
$|\Delta \bm{z}^\pm_\lambda|$ averaged over the direction
of~$\bm{\hat{s}}$, and we define $\theta_\lambda$ to be the (positive
semi-definite) angle between $\Delta \bm{z}^+_\lambda$ and $\Delta
\bm{z}^-_\lambda$ averaged over the direction of~$\bm{\hat{s}}$.  We
think of $\delta z^\pm_\lambda(\bm{x},t)$ as the characteristic
amplitude of the $z^\pm$ structure of scale~$\lambda$ that is located
at position~$\bm{x}$. Nonlinear interactions cause each structure at
scale~$\lambda$ to break up into a number of structures at smaller scales.
These smaller structures in turn break up into even
smaller structures, and so on.

As can be seen from Equation~(\ref{eq:RMHD}),
$\bm{z}^\pm$~fluctuations propagate with velocity~$\mp\bm{v}_{\rm
  A}$. We can thus view $z^-$ ($z^+$) structures as wave packets that
propagate parallel (anti-parallel) to the background magnetic field
while being distorted by nonlinear interactions.  The form of the
nonlinear term in Equation~(\ref{eq:RMHD}) implies
that nonlinear interactions occur only between
$z^+$ fluctuations and~$z^-$ fluctuations, and not between
fluctuations that propagate in the same
direction~\citep{iroshnikov63,kraichnan65}. The energy cascade in RMHD
turbulence can thus be viewed as resulting from ``collisions'' between
counter-propagating wave packets.  In the discussion below, we use the
terms ``structure,'' ``wave packet,'' and ``fluctuation''
interchangeably.

In the Appendix, we argue that if a $\delta z^\pm_\lambda$ fluctuation
collides with a $\delta z^\mp_\lambda$ fluctuation that is either much
stronger or much weaker than $\delta z^\pm_\lambda$, then $\lambda$
changes for both fluctuations (that is, they are sheared by each
other), but the fluctuation amplitudes remain approximately the
same. We refer to such collisions as ``highly imbalanced.'' On the
other hand, if $\delta z^+_\lambda \sim \delta z^-_\lambda$
(``balanced collisions''), then in general both $\lambda$ and the
fluctuation amplitudes decrease, as in models of non-intermittent MHD
and RMHD turbulence~\citep[e.g.,][]{goldreich95}.

To construct an analytic model of RMHD turbulence, we assume
that each balanced collision reduces a fluctuation's
amplitude by a constant factor~$\beta$, which satisfies
\begin{equation}
0 < \beta < 1,
\label{eq:betarange} 
\end{equation} 
 while highly imbalanced collisions
reduce~$\lambda$ without reducing a fluctuation's amplitude.   Thus,
\begin{equation}
\delta z^\pm_{\lambda} = \overline{ \delta z}\beta^q ,
\label{eq:z1} 
\end{equation} 
where $\overline{ \delta z}$ is the amplitude of the fluctuation's
``progenitor'' structure at the outer scale (or forcing scale)~$L$,
and $q$ is the number of balanced collisions experienced by the
fluctuation during its evolution from scale~$L$ to scale~$\lambda$.
For simplicity, we set\footnote{More realistically, $\overline{ \delta
    z}$ would have its own (scale-independent) distribution reflecting
  the non-universal details of the outer-scale statistics (e.g., the
  statistics of the forcing).}
\begin{equation}
\overline{ \delta z} = \mbox{ constant}.
\label{eq:zconst}  
\end{equation} 

To determine a plausible functional form for the probability
distribution function (PDF) of~$q$, we consider a hypothetical
scenario in which balanced collisions have the property that they
reduce a fluctuation's amplitude without changing its length scale.
In this case, balanced collisions are similar to the ``modulation
defect events'' described by \cite{she95}, in that a fluctuation's
amplitude can be reduced by a finite factor~$\beta$ during an interval
of time in which $\lambda$ decreases by only an infinitesimal
amount. If the length scale of a fluctuation decreases from~$L$
to~$\lambda$, then we can divide the interval $[0,\ln(L/\lambda)]$
into infinitesimal sub-intervals, and within each sub-interval there
is an infinitesimal chance that a modulation defect event occurs. Over
the entire interval, however, the average number of modulation defect
events is finite. If we assume that the probability of a balanced
collision is independent of the number of balanced collisions that
have already occurred, then~$q$ has a Poisson distribution,
\begin{equation}
P(q) = \frac{e^{-\mu}\mu^q }{q!},
\label{eq:Pq} 
\end{equation} 
where~$\mu$ is the as-yet-unknown, scale-dependent, mean value of~$q$.
In RMHD turbulence, balanced collisions do in fact change~$\lambda$,
and the probability that a balanced collision occurs may depend
upon~$q$.  Thus, the above arguments do not provide a rigorous
justification for Equation~(\ref{eq:Pq}). We proceed, however, using
Equation~(\ref{eq:Pq}) as a model.  We further assume that $\mu$ and
$\overline{ \delta z}$ are the same for $\delta z^+_\lambda$ and
$\delta z^-_\lambda$ and thereby restrict our analysis to the case of
zero cross helicity.

The median value of~$q$ is approximately~$\mu$~\citep{choi94}, and
thus the ``typical'' value of~$\delta z^\pm_\lambda$ that best
characterizes the bulk of the volume is
\begin{equation}
\delta z^\ast_\lambda = \overline{\delta z} \beta^\mu.
\label{eq:zast} 
\end{equation} 
In contrast, the most intense structures at scale~$\lambda$ correspond
to~$q=0$ and occur with probability~$e^{-\mu}$.
Equation~(\ref{eq:z1}) implies that the variation in~$\bm{z}^+$ or
$\bm{z}^-$ across such a $q=0$ structure is~$\overline{\delta z}$,
independent of~$\lambda$. We assume that these structures correspond
to sheet-like quasi-discontinuities (current/vorticity sheets) with a
volume-filling factor~$\propto
\lambda$~\citep[c.f.,][]{grauer94,politano95}. Setting~$e^{-\mu}
\propto \lambda$, we obtain
\begin{equation}
\mu = A + \ln\left(\frac{L}{\lambda}\right),
\label{eq:chi1} 
\end{equation} 
where~$A$ is a constant that quantifies the breadth of the
distribution at the outer scale.
We can thus rewrite Equation~(\ref{eq:zast})  in the form
\begin{equation}
\delta z^\ast_\lambda = \overline{ \delta z}\left(\frac{\lambda}{e^A L}\right)^{-\ln\beta}.
\label{eq:zast2}
\end{equation} 

\vspace{0.0cm} 
\subsection{Timescales and Critical Balance}
\label{sec:timescales} 
\vspace{0.0cm} 

We define the nonlinear timescale
\begin{equation}
\tau_{{\rm nl},\lambda}^\pm = \frac{\lambda}{\delta z^\mp_\lambda \sin\theta_\lambda},
\label{eq:deftaunl} 
\end{equation} 
which is the rate at which a $z^\pm$ structure at scale~$\lambda$ is
sheared by the~$z^\mp$ structure at scale~$\lambda$ at that same location.
The factor of $\sin\theta_\lambda$ is included in
Equation~(\ref{eq:deftaunl}) because,
if $\bm{z}^+$ and $\bm{z}^-$ are aligned to
within a small angle~$\theta$, then $|\bm{z}^\mp \cdot \nabla
\bm{z}^\pm|$ is reduced by a factor~$\sim \theta$ relative to the case
in which $\theta\sim 1$~\citep{boldyrev05}.  

We define the linear timescale
\begin{equation}
\tau_{\rm lin,\lambda}^\pm = \frac{l^\pm_\lambda}{v_{\rm A}}.
\label{eq:tau_lin} 
\end{equation} 
Here, $l^\pm_\lambda$ is the ``parallel'' correlation length of a
$\delta z^\pm_\lambda$ structure measured along a local mean magnetic
field, which is obtained by summing~$\bm{B}_0$ with all the magnetic
fluctuations at scales that exceed~$\lambda$ by a factor of at least
a~few.

In accord with the critical-balance hypothesis of
\cite{goldreich95}, we assume that
\begin{equation}
\tau_{{\rm nl},\lambda}^\pm \sim \tau_{{\rm lin},\lambda}^\pm.
\label{eq:CB0} 
\end{equation} 
We can rewrite Equation~(\ref{eq:CB0})
as\footnote{Equation~(\ref{eq:CB}) is a simplifying assumption. In
  numerical simulations of RMHD turbulence, $\chi^\pm$ has a
  distribution, but this distribution is scale-independent and has a
  mean of order unity (Mallet et al.~2014, in preparation).}
\begin{equation}
 \chi^\pm = \frac{l^\pm_\lambda \delta z^\mp_\lambda
  \sin\theta_\lambda}{\lambda v_{\rm A}}\sim 1.
\label{eq:CB} 
\end{equation} 
Equation~(\ref{eq:CB0}) is also equivalent to the relation
\begin{equation}
l^\pm_\lambda \sim v_{\rm A} \tau_{{\rm nl},\lambda}^\pm,
\label{eq:lpmval} 
\end{equation} 
which states that the parallel correlation length
of a $\delta z^\pm_\lambda$ fluctuation is roughly the distance it can
propagate during its cascade timescale.  

When a $\delta z^+_\lambda$ fluctuation collides with a $\delta
z^-_\lambda$ fluctuation and $\delta z^+_\lambda \sim \delta
z^-_\lambda \sim \delta z^\ast_\lambda$ (see
Equation~(\ref{eq:zast})), nonlinear interactions cause the $\delta
z^-_\lambda$ and $\delta z^+_\lambda$ fluctuations to evolve on the
same timescale in a strongly coupled and unpredictable way, which
impedes the development of alignment.  We thus take
\begin{equation}
\left.\begin{array}{c}
\displaystyle \theta_\lambda \sim 1 \vspace{0.3cm} \\
\displaystyle \tau_{{\rm nl},\lambda}^\pm \sim \lambda/\delta z^\ast_\lambda 
\end{array}\hspace{0.3cm} 
\right\}  \hspace{0.1cm} \mbox{(when $\delta z^+_\lambda \sim \delta z^-_\lambda \sim \delta z^\ast_\lambda$)}.
\label{eq:balanced} 
\end{equation} 
The characteristic parallel correlation length of the
median-amplitude fluctuations at scale~$\lambda$ is then
\begin{equation}
l^\ast_\lambda = \frac{v_{\rm A} \lambda }{\delta z^\ast_\lambda}.
\label{eq:last} 
\end{equation}

\vspace{0.0cm} 
\subsection{Nonlinear Interactions and the Refined Similarity Hypothesis}
\label{sec:NL} 
\vspace{0.0cm} 

Given our assumption that~$\tau_{{\rm nl},\lambda}^\pm \sim \tau_{{\rm
    lin},\lambda}^\pm$, the turbulence is strong, and $\delta
z^\pm_\lambda$ energy cascades to smaller scales on the
timescale~$\tau_{{\rm nl},\lambda}^\pm$.  We define
$\epsilon^\pm_\lambda$ to be the rate at which $z^\pm$ energy (per
unit mass) is dissipated within a sphere of diameter~$\lambda$.  In
keeping with earlier works on intermittency, we take~$\epsilon^\pm$ to
be ``equal in law'' to the quantity $(\delta
z^\pm_\lambda)^2/\tau_{{\rm nl},\lambda}^\pm$~\citep{frisch96}.  This means that the
$n^{\rm th}$ moment of~$\epsilon^\pm_\lambda$ and the $n^{\rm th}$
moment of the quantity $(\delta z^\pm_\lambda)^2/\tau_{{\rm
    nl},\lambda}^\pm$ scale with~$\lambda$ in the same way for
all~$n$.  We denote ``equality in law'' with the symbol~$\approx$ and
thus write
\begin{equation}
\epsilon_\lambda^\pm \approx \frac{(\delta z^\pm_\lambda)^2 \delta z^\mp_\lambda
  \sin\theta_\lambda}{\lambda}.
\label{eq:eps1} 
\end{equation} 
Equation~(\ref{eq:eps1}) is analogous to Kolmogorov's~(1962) refined
similarity hypothesis for hydrodynamic
turbulence.\nocite{kolmogorov62} We note that the scalings that we
derive below do not require full ``equality in law,'' but just that
the averages of the left- and right-hand sides of
Equation~(\ref{eq:eps1}) scale with~$\lambda$ in the same way. 

The average dissipation rate within a sphere of diameter~$\lambda$
is independent of~$\lambda$ and thus satisfies the relation
\begin{equation}
\langle \epsilon^\pm_\lambda \rangle = \epsilon,
\label{eq:exactlaw} 
\end{equation} 
where $\langle \dots \rangle$ indicates a spatial average
and~$\epsilon$ is the global, average dissipation rate, which is the
same for~$z^+$ and~$z^-$ fluctuations given our assumption that the
cross helicity is zero. In forced turbulence that has reached a
(statistical) steady state, $\epsilon$ is also the rate at which
energy is injected into the turbulence at the outer scale.  Our goal
now is to use Equations~(\ref{eq:eps1}) and~(\ref{eq:exactlaw}) to
determine the value of~$\beta$.

\begin{figure}[t]
\centerline{
\includegraphics[width=8cm]{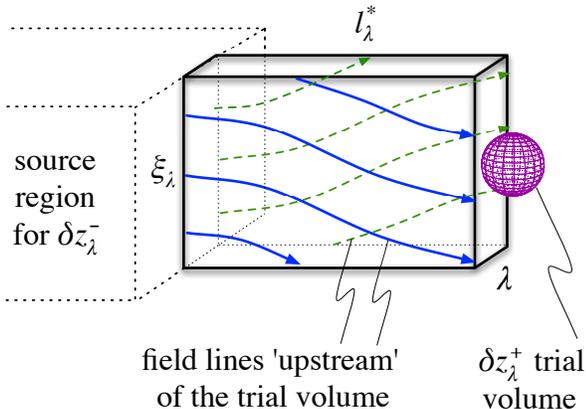}
}
\caption{Our analysis focuses on the $z^+$ cascade power~$(\delta
  z^+_\lambda)^2/\tau_{{\rm nl},\lambda}^+$ within a spherical trial
  volume of diameter~$\lambda$ in which~$\delta z^+_\lambda \gg \delta
  z^\ast_\lambda$ and $\delta z^+_\lambda \gg \delta z^-_\lambda$. The
  intense $\delta z^+_\lambda$ fluctuation in the trial volume is part
  of a sheet-like coherent structure of thickness~$\lambda$. The
  $\delta z^-_\lambda$ fluctuation within the trial volume can be
  viewed as originating from a source region a distance~$\sim
  l^\ast_\lambda$ from the trial volume ``upstream'' along the
  magnetic field.  Field lines on opposite sides of the~$\delta
  z^+_\lambda$ structure that are initially a distance~$\lambda$ apart
  separate by a distance~$ \sim \xi_\lambda = l^\ast_\lambda \delta
  z^+_\lambda/v_{\rm A}$ when they are followed for a
  distance~$l^\ast_\lambda$.
\vspace{0.2cm} 
\label{fig:source_region} }
\end{figure}

To average the right-hand side of Equation~(\ref{eq:eps1}), we consider the
spherical trial volume of diameter~$\displaystyle \lambda$ illustrated in
Figure~\ref{fig:source_region}.  We can take the PDF of~$\delta
z^+_\lambda$ within the trial volume to be determined by
Equations~(\ref{eq:z1}) and (\ref{eq:Pq}). However, once we do so, we
cannot also take the value of~$\delta z^-_\lambda$ within the trial
volume to have the same distribution, because in general $\delta
z^+_\lambda$ and $\delta z^-_\lambda$ are correlated.

We assume that the dominant contribution to~$\langle (\delta
z^+_\lambda)^2 \delta z^-_\lambda \sin \theta_\lambda/\lambda \rangle$
comes from exceptionally intense~$\delta z^+_\lambda$ fluctuations
satisfying the inequalities $\displaystyle \delta z^+_\lambda \gg
\delta z^\ast_\lambda$ and $\displaystyle \delta z^+_\lambda \gg
\delta z^-_\lambda$.  We therefore focus on the case in which these
inequalities are satisfied within the trial volume.  We assume (and
confirm below in Equation~(\ref{eq:lplus2}))
that $l^+_\lambda \gg l^\ast_\lambda$ when $\delta z^+_\lambda \gg
\delta z^\ast_\lambda$.  Because of their comparatively large parallel
correlation lengths (and long lifetimes, as we will see in
Equation~(\ref{eq:taunl1})), we refer to structures with $\delta
z^+_\lambda \gg \delta z^\ast_\lambda$ as coherent structures.

The $z^\pm$ fluctuations at scale~$\lambda$ propagate along a local
magnetic field obtained by summing~$\bm{B}_0$ with the magnetic-field
fluctuations at length scales exceeding~$\lambda$ by some factor of
order unity. Because this factor is not uniquely determined, the
direction in which $\delta z^\pm_\lambda$ fluctuations propagate is
only determined to within an angular uncertainty of order
\begin{equation}
\Delta \theta_\lambda \sim \frac{\delta z^\ast_\lambda}{v_{\rm A}},
\label{eq:Dtheta} 
\end{equation} 
where we have taken the fluctuations at scales somewhat larger
than~$\lambda$ to have amplitudes comparable to~$\delta
z^\ast_\lambda$. Here we have assumed that the intense~$\delta
z^+_\lambda$ fluctuations propagate through a background of
median-amplitude $z^-$ fluctuations, a point that we discuss further
in connection with Equation~(\ref{eq:dzmxiast2}) below.  Because of
the angular uncertainty~$\Delta \theta_\lambda$, a $z^-$ structure of
scale~$\lambda$ is only able to propagate a distance~$\displaystyle
\sim \lambda/(\Delta \theta_\lambda) \sim l^\ast_\lambda$ (see
Equation~(\ref{eq:last})) through a counter-propagating $\displaystyle
\delta z^+_\lambda$ structure before propagating out of that
structure.

Because of this, if we follow the magnetic-field lines in the trial
volume (Figure~\ref{fig:source_region}) back along the magnetic field
a distance~$l^\ast_\lambda$, we reach a ``source region'' in which the
$z^-$ fluctuations have not yet interacted with the coherent $\delta
z^+_\lambda$ structure. Within this source region, the $z^-$
fluctuations are not aligned with the coherent $\delta z^+_\lambda$
structure, because they do not yet ``know about'' the coherent $\delta
z^+_\lambda$ structure's orientation in space.  
If we pick two field lines a distance~$\lambda$ apart
within the trial volume and follow them for a
distance~$l^\ast_\lambda$, they will typically separate by a distance
of order
\begin{equation}
\xi_\lambda =   \frac{l^\ast_\lambda \delta z^+_\lambda}{v_{\rm A}} = 
\lambda\,\frac{ \delta z^+_\lambda}{\delta z^\ast_\lambda} \gg \lambda.
\label{eq:xiast} 
\end{equation} 
We assume that the coherent~$\delta z^+_\lambda$ structure remains
coherent over a distance of at least~$\sim \xi_\lambda$ in the direction of
the vector magnetic-field fluctuations associated with the $\delta
z^+_\lambda$ structure --- i.e., throughout the slab depicted in
Figure~\ref{fig:source_region}. The coherent~$\delta z^+_\lambda$
structure is thus sheet-like.

We expect (and confirm below in Equation~(\ref{eq:taunl1})) that the
cascade timescale of the coherent $\delta z^+_\lambda$ structure
is~$\gg l^\ast_\lambda/v_{\rm A}$, so that the $\delta z^+_\lambda$
structure changes very little as a $z^-$ fluctuation propagates from
the source region to the trial volume.  We make the approximation that
during this transit, the $z^-$ fluctuation evolves as if it were acted
upon by a linear~$z^+$ shear with shearing rate~$\delta
z^+_\lambda/\lambda$ that lasts for a time~$l^\ast_\lambda/v_{\rm A}$,
where the term ``linear'' refers to the shear's spatial
profile (see Equation~(\ref{eq:zplusshear})).  In the Appendix,
we present an analytic calculation showing that in this approximation
the amplitude of the $z^-$ fluctuation is unchanged by the shear, but
the $z^-$ fluctuation is rotated into alignment so that
\begin{equation}
\sin\theta_\lambda \simeq \theta_\lambda \sim
\frac{\lambda}{\xi_\lambda} = \frac{\delta z^\ast_\lambda}{\delta
  z^+_\lambda}
\label{eq:thetapm0} 
\end{equation} 
within the trial volume.  We also show in the Appendix that, because
the $z^-$ fluctuations are sheared at rate~$\delta
z^+_\lambda/\lambda$ for a time~$l^\ast_\lambda/v_{\rm A}$, their
perpendicular scales decrease by a factor of
\begin{equation}
\frac{\delta z^+_\lambda}{\lambda} \times \frac{l^\ast_\lambda}{v_{\rm A}}
= \frac{\xi_\lambda}{\lambda} \gg 1
\label{eq:scale_reduction} 
\end{equation} 
during their propagation from the source region to the trial volume.
The source region in Figure~\ref{fig:source_region} contains~$z^-$
fluctuations spanning a range of perpendicular scales. According to
the above arguments, the fluctuations at scale~$\xi_\lambda$ in the
source region make the dominant contribution to the values of~$\delta
z^-_\lambda$ and $\delta z^-_\lambda \sin\theta_\lambda$ within the
trial volume.  Thus,
\begin{equation}
\delta z^-_\lambda \bigg|_{\rm trial\;volume} \simeq \;\delta z^-_{\xi_\lambda}\bigg|_{\rm source\;region}.
\label{eq:tvsr1} 
\end{equation} 

\begin{figure}[t]
\centerline{
\includegraphics[width=8cm]{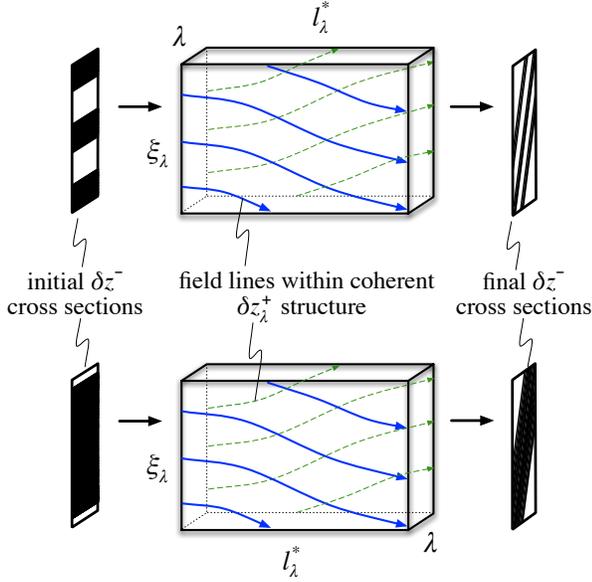}
}
\caption{As $z^-$ fluctuations propagate a distance~$l^\ast_\lambda$
  through the coherent~$\delta z^+_\lambda$ structure, the pattern of
  the~$z^-$ fluctuations in the field-perpendicular plane
  approximately follows the magnetic field lines within the coherent
  $\delta z^+_\lambda$ structure. This causes the perpendicular length
  scale of the~$z^-$ fluctuations to decrease by a factor~$\sim
  \xi_\lambda/\lambda \gg 1$ and rotates the $z^-$ fluctuations into
  alignment with the $\delta z^+_\lambda$ structure, decreasing the
  angle between the $\bm{z}^+$ and~$\bm{z}^-$ fluctuations to a
  value~$ \sim \lambda/\xi_\lambda$.
\vspace{0.2cm} 
\label{fig:shear_slab} }
\end{figure}

The top half of Figure~\ref{fig:shear_slab} illustrates the arguments
underlying Equation~(\ref{eq:thetapm0}) and the scale-reduction factor in
Equation~(\ref{eq:scale_reduction}) for the hypothetical case in which
the~$z^-$ fluctuations in the source region have square cross sections
of scale~$\lambda$ in the field-perpendicular plane. In the trial
volume, the perpendicular scale length of these fluctuations
becomes~$\sim \lambda^2 /\xi_\lambda$ and, because of
Equation~(\ref{eq:thetapm0}), $\theta_\lambda \ll 1$. The evolution
of~$z^-$ can be recovered heuristically by taking the pattern of the
$z^-$ fluctuation in the field-perpendicular plane to follow the
perturbed magnetic field lines and by taking the direction of the
$\bm{z}^-$ fluctuation in the trial volume to become approximately
parallel to the striated $z^-$~pattern, so as to preserve the
incompressibility condition. The bottom half of
Figure~\ref{fig:shear_slab} illustrates the evolution of a $z^-$ eddy
of perpendicular length scale~$\sim \xi_\lambda$.  Within the trial
volume, the perpendicular length scale of this fluctuation
becomes~$\sim \lambda$.

Equations~(\ref{eq:thetapm0}) and (\ref{eq:tvsr1}) 
imply that Equation~(\ref{eq:eps1})  becomes
\begin{equation}
\epsilon^+_\lambda \approx \frac{ (\delta z^+_\lambda)^2 \delta z^-_{\xi_\lambda}}{\xi_\lambda}
\label{eq:eps2} 
\end{equation} 
when $\delta z^+_\lambda \gg \delta z^\ast_\lambda$, where
$\epsilon^+_\lambda$ and $\delta z^+_\lambda$ are evaluated within the
trial volume and $\delta z^-_{\xi_\lambda}$ is evaluated within the
source region.  We now consider the average of
Equation~(\ref{eq:eps2}). Our assumption that $\delta z^+_\lambda
\gg \delta z^\ast_\lambda$ in the trial volume decreases the
probability that $\delta z^-_{\xi_\lambda}$ is much larger than
$\delta z^\ast_{\xi_\lambda}$ in the source region, because intense,
counter-propagating, $z^\pm$ structures rapidly annihilate. We thus
take the PDF of~$\delta z^-_{\xi_\lambda}$ within the source region to
be negligible at large $\delta z^-_{\xi_\lambda}$ and make the
approximation that
\begin{equation}
\delta z^-_{\xi_\lambda}\bigg|_{\rm source\;region} \simeq \delta z^\ast_{\xi_\lambda}.
\label{eq:dzmxiast2} 
\end{equation} 
It follows from Equation~(\ref{eq:zast2})  that
$\delta z^\ast_{\xi_\lambda}/\xi_\lambda =(\delta z^\ast_\lambda/\lambda)
(\xi_\lambda/\lambda)^{-1 -\ln \beta}$, and thus, using
Equation~(\ref{eq:xiast}), we can rewrite Equation~(\ref{eq:eps2}) as
\begin{equation}
\epsilon^+_\lambda \approx \frac{(\delta z^+_\lambda)^{1-\ln\beta} (\delta z^\ast_\lambda)^{2+\ln\beta}}{\lambda}.
\label{eq:eps3}  
\end{equation} 
We now average Equation~(\ref{eq:eps3}) over space. For the right-hand
side of Equation~(\ref{eq:eps3}), this is equivalent to averaging
over the Poisson distribution of~$q$ in 
Equation~(\ref{eq:z1}), which is given in Equation~(\ref{eq:Pq}). 
We thus obtain
 \begin{equation}
\langle \epsilon^+_\lambda \rangle \sim \frac{(\overline{ \delta z})^3}{\lambda}\,
\beta^{\mu(2+\ln\beta)} e^{-\mu} \sum_{q=0}^\infty \frac{(\mu \beta^{1-\ln\beta})^q}{q!}.
\label{eq:eps4} 
\end{equation} 
To derive Equation~(\ref{eq:eps3}), we assumed that $\delta
z^+_\lambda \gg \delta z^\ast_\lambda$, and as a consequence the form
of the summand in Equation~(\ref{eq:eps4}) is incorrect when~$q\gtrsim
\mu$. However, in the inertial range, in which~$\mu$ is formally
large, terms with $q\gtrsim \mu$ make only a small contribution to the
sum in Equation~(\ref{eq:eps4}),\footnote{We verify this claim by
  first writing the sum in Equation~(\ref{eq:eps4}) as $I \equiv
  \sum_{q=0}^\infty (\alpha \mu)^q/q! = e^{\alpha \mu}$, where $\alpha
  = \beta^{1-\ln\beta} < 1$.  For simplicity, we take~$\mu$ to be an
  integer.  The contribution to~$I$ from terms with $q> \mu$ (which we
  denote~$I_1$) then equals $e^{\alpha\mu} \gamma(\mu+1,\alpha
  \mu)/\mu!$, where $\gamma$ is the lower incomplete gamma
  function. When $a$ exceeds $x$ by a factor of order unity and $x\gg
  1$, $\gamma(a+1,x) \simeq e^{-x}x^{a+1}/(a-x)$ \citep{ferreira05}.
  With the use of Stirling's formula, $\mu! \simeq
  \sqrt{2\pi\mu}(\mu/e)^\mu$, we obtain $I_1 \simeq
       [\alpha/(1-\alpha)](2\pi\mu)^{-1/2}\exp(\mu(1+\ln\alpha))$,
       which is~$\ll I$ because $1 + \ln\alpha < \alpha$.}  consistent
with our assumption that the total $z^+$ dissipation rate is dominated
by large-$\delta z^+_\lambda$ regions.  The sum in
Equation~(\ref{eq:eps4}) is simply $\exp\left(\mu
\beta^{1-\ln\beta}\right)$, and thus Equation~(\ref{eq:eps4}) implies
that
\begin{equation}
\langle \epsilon^+_\lambda \rangle  \propto \left(
\frac{\lambda}{L}\right)^{-(2+\ln\beta)\ln \beta - \beta^{1-\ln\beta}}.
\label{eq:eps5} 
\end{equation} 
Since $\langle \epsilon^+_\lambda \rangle$ must be independent
of~$\lambda$, we obtain
\begin{equation}
(2+\ln\beta) \ln \beta  + \beta^{1-\ln\beta} = 0.
\label{eq:beta0} 
\end{equation} 

There are two solutions to Equation~(\ref{eq:beta0}): $\beta \simeq
0.136$ and $\beta \simeq 0.691$. The solution $\beta \simeq 0.136$
leads to the scaling $\delta z^\ast_\lambda \propto \lambda^{1.98}$,
which implies that the median variation in~$\bm{z}^\pm$ across a
distance~$\lambda$ measured perpendicular to~$\bm{B}$ is dominated by
the outer-scale eddies and not by~$\delta z^\ast_\lambda$, as we have
assumed.  Thus, the only solution to Equation~(\ref{eq:beta0}) that is
consistent with its derivation is
\begin{equation}
\beta \simeq 0.691.
\label{eq:beta1} 
\end{equation} 
We note that Equations~(\ref{eq:exactlaw}), (\ref{eq:eps4}),
and~(\ref{eq:beta0}) imply that
\begin{equation}
\epsilon \sim \frac{(\overline{ \delta z})^3}{e^A L}.
\label{eq:eps6} 
\end{equation} 
Equation~(\ref{eq:eps6}) establishes a relationship between the energy
input into the turbulence and the two parameters $\overline{ \delta
  z}$ and~$A$ that quantify the non-universal features of the
outer-scale fluctuations. Given~$\epsilon$ and~$L$, only one
of~$\overline{ \delta z}$ and~$A$ is a free parameter in our model.

\vspace{0.0cm} 
\subsection{Consistency of the Strong-Turbulence Assumption}
\label{sec:strong} 
\vspace{0.0cm} 

As described in Section~\ref{sec:NL}, the type of nonlinear
interaction that is most effective at shearing large-amplitude $\delta
z^+_\lambda$ structures involves the typical $\delta z^-_{\xi_\lambda}
\sim \delta z^\ast_{\xi_\lambda}$ structures
(Equation~(\ref{eq:dzmxiast2})) at scale~$\xi_\lambda$, which
exceeds~$\lambda$ to a degree that depends on the amplitude $\delta
z^+_\lambda$ (see Equation~(\ref{eq:xiast})).  These are the
$z^-$~structures in the source region that, upon shearing by an
intense $\delta z^+_\lambda$ structure, become the $\delta
z^-_\lambda$ structures in the trial volume in
Figure~\ref{fig:source_region} (see Equation~(\ref{eq:tvsr1})) and
enter into the computation of the average cascade power $\langle
(\delta z^+_\lambda)^2 \delta z^-_\lambda \sin \theta_\lambda/\lambda
\rangle$ within the trial volume.  The parallel correlation length of
the typical $\delta z^\ast_{\xi_\lambda}$ fluctuations in the source
region in Figure~\ref{fig:source_region} is~$\sim
l^\ast_{\xi_\lambda}$, and thus the correlation time of the $\delta
z^-_\lambda$ fluctuation in the trial volume is~$\sim
l^\ast_{\xi_\lambda}/v_{\rm A}$ (assuming that the $\delta
z^+_\lambda$ structure does not decorrelate on a shorter timescale, as
we now demonstrate).  Using Equations~(\ref{eq:thetapm0}),
(\ref{eq:tvsr1}), and~(\ref{eq:dzmxiast2}), we rewrite
Equation~(\ref{eq:deftaunl}) in the form
\begin{equation}
\tau_{{\rm nl},\lambda}^+ \sim \frac{\xi_\lambda}{\delta
  z^\ast_{\xi_\lambda}} = \frac{l^\ast_{\xi_\lambda}}{v_{\rm A}},
\label{eq:tau0} 
\end{equation} 
where we have used Equation~(\ref{eq:last}) to deduce that
$\xi_\lambda/\delta z^\ast_{\xi_\lambda} =l^\ast_{\xi_\lambda}/v_{\rm
  A}$.  Thus, the cascade timescale~$\tau_{{\rm nl},\lambda}^+$ of a
large-amplitude, coherent, $\delta z^+_\lambda$ structure is also the
correlation timescale of the $z^-$ fluctuations that dominate the
shearing of that $\delta z^+_\lambda$ structure. This consistency
check confirms that the turbulence is strong, as we have assumed.

\vspace{0.0cm} 
\subsection{Locality}
\label{sec:locality} 
\vspace{0.0cm} 

In Equation~(\ref{eq:eps1}), we assumed that the cascade is local
in~$\lambda$, in the sense that $z^\pm$ structures are sheared
primarily by the counter-propagating $z^\mp$ structures of similar
perpendicular scale at the same location. On the other hand, as we
have just summarized in Section~\ref{sec:strong}, an intense $\delta
z^+_\lambda$ fluctuation is cascaded primarily by collisions with
$z^-$ fluctuations whose perpendicular scale prior to colliding
was~$\xi_\lambda$, which significantly exceeds~$\lambda$. Thus, the
cascade is local in~$\lambda$ if the scales of the interacting
fluctuations are evaluated at the same point in space (e.g., the trial
volume in Figure~\ref{fig:source_region}), but nonlocal if the
perpendicular scale of~$\delta z^+_\lambda$ is evaluated in the trial
volume in Figure~\ref{fig:source_region} while the perpendicular scale
of the $z^-$ fluctuation is evaluated in the source region depicted in
this figure. We note that Equations~(\ref{eq:lpmval}) and
(\ref{eq:tau0}) imply that, when~$\delta z^+_\lambda \gg \delta
z^\ast_\lambda$,
\begin{equation}
l^+_\lambda \sim l^\ast_{\xi_\lambda}.
\label{eq:lpmval2} 
\end{equation} 
Thus, just before the nonlinear interaction begins, the $z^-$
fluctuations that dominate the shearing of a large-amplitude,
coherent~$\delta z^+_\lambda$ structure have the same parallel
correlation length as that~$\delta z^+_\lambda$ structure.  In this
sense, the cascade could be described as ``local in parallel length
scale.'' 

\vspace{0.0cm} 
\subsection{Inertial-Range Scalings}
\label{sec:scalings} 
\vspace{0.0cm} 

The two-point structure functions $\langle (\delta z^\pm_\lambda)^n\rangle$
are the standard measures used to establish
the presence of intermittency in
turbulence~\citep{kolmogorov62,frisch96}.  From Equations
(\ref{eq:z1}) through~(\ref{eq:chi1}), we obtain
\begin{equation}
\langle (\delta z^\pm_\lambda)^n\rangle = (\overline{\delta z})^{n} 
e^{-\mu} \sum_{q=0}^{\infty} \frac{\left(\mu \beta^n\right)^q}{q!}.
\label{eq:sf0} 
\end{equation} 
The sum in Equation~(\ref{eq:sf0}) is
simply~$e^{\mu\beta^n}$. With the use of Equation~(\ref{eq:chi1}), we
thus obtain
\begin{equation}
\langle (\delta z^\pm_\lambda)^n\rangle = (\overline{\delta z})^{n} 
\left(\frac{\lambda}{e^A L}\right)^{\zeta_n},
\label{eq:sfs} 
\end{equation} 
where 
\begin{equation} 
\zeta_n = 1 - \beta^n .
\label{eq:xin} 
\end{equation} 
The summand in Equation~(\ref{eq:sf0})  is maximized when $q\simeq q_n$, where
\begin{equation}
q_n = \mu\beta^{n}.
\label{eq:qn} 
\end{equation} 
Terms with $q< q_n$ account for approximately half of the total sum in
Equation~(\ref{eq:sf0}), just as the median of~$P(q)$ in
Equation~(\ref{eq:Pq}) is approximately~$\mu$~\citep{choi94}. 
The mean value of~$q$ is~$\mu$, and the standard deviation of~$q$ is
\begin{equation}
\sigma = \langle (q - \mu)^2 \rangle^{1/2} = \mu^{1/2}.
\label{eq:qrms} 
\end{equation} 
Thus, the fluctuations that make the dominant contribution to $\langle
(\delta z^\pm_\lambda)^n\rangle$ are~$\sim N$ standard deviations out
into the tail of the $q$~distribution, where
\begin{equation}
N = \frac{\mu-q_n}{\sigma} = \mu^{1/2}(1-\beta^n).
\label{eq:qnsigma} 
\end{equation} 
As~$\lambda$ decreases, $N$ increases, and this increase is more rapid
when~$n$ is larger. It is this fact that allows $\langle (\delta
z^\pm_\lambda)^n\rangle$ in Equation~(\ref{eq:sfs}) to decrease more
slowly with decreasing~$\lambda$ than does $\langle \delta
z^\pm_\lambda\rangle^n$.

From Equations~(\ref{eq:sfs}) and (\ref{eq:xin}),
the second-order structure function satisfies the relation
\begin{equation}
\langle (\delta
z_\lambda^+)^2\rangle \propto \lambda^{1 - \beta^2} \simeq \lambda^{0.52}, 
\label{eq:zsqd} 
\end{equation} 
which corresponds to an inertial-range $z^\pm$ power
spectrum
\begin{equation}
E(k_\perp) \propto k_\perp^{-1.52},
\label{eq:Ekperp} 
\end{equation} 
where $k_\perp$ is the wave-vector component perpendicular to~$\bm{B}_0$.
Equation~(\ref{eq:sfs}) implies that the kurtosis obeys the scaling
\begin{equation}
\frac{\langle (\delta z_\lambda^+)^4 \rangle}{\langle (\delta z_\lambda^+)^2\rangle ^2 }
= \left(\frac{ \lambda}{e^A L}\right)^{- (1 -\beta^2)^2} \propto \lambda^{-0.27},
\label{eq:kurtosis} 
\end{equation} 
which exemplifies how intermittency increases with
decreasing~$\lambda$.  We emphasize that the parameter~$A$ does not
affect the exponents in any of the power-law scalings in our model
(nor the fact that $\theta^\star_\lambda$ in
Equation~(\ref{eq:thetastarlim}) below decreases logarithmically
as~$\lambda/L $ decreases to very small values).

For reference, Equation~(\ref{eq:zast2}) implies that the amplitude of
a ``typical'' structure is
\begin{equation}
\delta z^\ast_\lambda = 
 \overline{ \delta z}\left(\frac{\lambda}{e^A L}\right)^{-\ln\beta}
\propto  \lambda^{0.37},
\label{eq:zastsc} 
\end{equation} 
and hence
\begin{equation}
\frac{\langle (\delta z^+_\lambda)^2 \rangle^{1/2}}{\delta
  z^\ast_\lambda} =\left(\frac{\lambda}{e^A L}\right)^{\ln\beta +
  (1-\beta^2)/2} \propto \lambda^{-0.11}.
\label{eq:mean_median} 
\end{equation} 
This shows that at
small~$\lambda/L$
the rms fluctuation
amplitude is much larger than the median fluctuation amplitude. Equation~(\ref{eq:zastsc}) implies via Equation~(\ref{eq:last}) that
\begin{equation}
l^\ast_\lambda \propto \lambda^{1+\ln\beta} \simeq \lambda^{0.63}.
\end{equation}
Equations (\ref{eq:thetapm0}), (\ref{eq:tvsr1})
and~(\ref{eq:dzmxiast2}) imply that, when $\delta z^+_\lambda \gg
\delta z^\ast_\lambda$,
\begin{equation}
\tau_{{\rm nl},\lambda}^+ \sim \frac{\lambda}{\delta z^\ast_\lambda} \left(
\frac{\delta z^+_\lambda}{\delta z^\ast_\lambda}\right)^{1 + \ln\beta}.
\label{eq:taunl1} 
\end{equation} 
The energy cascade timescale of the most
intense fluctuations $\tau_{\rm max}$ follows from setting~$\delta
z^+_\lambda = \overline{\delta z}$ in Equation~(\ref{eq:taunl1}),
which, together with Equation~(\ref{eq:zastsc}), yields
\begin{equation}
\tau_{\rm max} \propto \lambda^{(1 +\ln \beta)^2} \simeq
\lambda^{0.40}.
\label{eq:taumax} 
\end{equation} 
Finally, Equations~(\ref{eq:lpmval}) and (\ref{eq:taunl1}) yield
\begin{equation}
l^+_\lambda \sim l^\ast_\lambda \left(\frac{\delta z^+_\lambda}{\delta z^\ast_\lambda}\right)^{1+\ln\beta} \gg l^\ast_\lambda
\label{eq:lplus2} 
\end{equation} 
(which confirms an assumption to this effect in Section~\ref{sec:NL}).

\vspace{0.0cm} 
\subsection{Alignment}
\label{sec:alignment} 
\vspace{0.0cm} 

We define the average
alignment angles
\begin{equation}
\theta^\pm_\lambda = \frac{\langle |\Delta \bm{z}^+_\lambda \times \Delta \bm{z}^-_\lambda|\rangle}{
\langle| \Delta \bm{z}^+_\lambda|| \Delta \bm{z}^-_\lambda|\rangle}
\label{eq:thetapm} 
\end{equation} 
and
\begin{equation}
\theta^{(vb)}_\lambda = \frac{\langle |\Delta \bm{v}_\lambda \times\Delta \bm{b}_\lambda|\rangle}{
\langle |\Delta \bm{v}_\lambda| |\Delta \bm{b}_\lambda|\rangle},
\label{eq:thetavb} 
\end{equation} 
where $\Delta \bm{v}_\lambda = (\Delta \bm{z}^+_\lambda + \Delta \bm{z}^-_\lambda)/2$,
$\Delta \bm{b}_\lambda = (\Delta \bm{z}^+_\lambda - \Delta \bm{z}^-_\lambda)/2$,
and $\langle \dots \rangle$ now denotes
averages over volume as well as the direction of the unit
vector~$\bm{\hat{s}}$ defined following Equation~(\ref{eq:Deltaz}).
To evaluate $\theta^\pm_\lambda$, we set
\begin{equation}
\langle |\Delta \bm{z}^+_\lambda \times \Delta \bm{z}^-_\lambda |\rangle
\sim \langle \delta z^+_\lambda \delta z^-_\lambda \sin\theta_\lambda \rangle.
\label{eq:approx1} 
\end{equation} 
We assume (and verify below) that in the inertial range the dominant
contribution to $ \langle \delta z^+_\lambda \delta z^-_\lambda
\sin\theta_\lambda \rangle$ comes from regions in which $\delta
z^+_\lambda \gg \delta z^\ast_\lambda$ or $\delta z^-_\lambda \gg
\delta z^\ast_\lambda$.  Since $ \langle \delta z^+_\lambda \delta
z^-_\lambda \sin\theta_\lambda \rangle$ is symmetric with respect to
the interchange of $z^+$ and $z^-$, we can estimate $ \langle \delta
z^+_\lambda \delta z^-_\lambda \sin\theta_\lambda \rangle$ by keeping
only the contribution from regions in which $\delta z^+_\lambda \gg
\delta z^\ast_\lambda$.  We then evaluate this contribution by
considering a spherical trial volume of diameter~$\lambda$ as in
Figure~\ref{fig:source_region} and approximating $\delta z^-_\lambda$
and $\sin\theta_\lambda$ within the trial volume using
Equations~(\ref{eq:thetapm0}), (\ref{eq:tvsr1}), and
(\ref{eq:dzmxiast2}).  We then average over the log-Poisson PDF
of~$\delta z^+_\lambda$ and make use of Equation~(\ref{eq:beta0}) to
obtain
\begin{equation}
\langle |\Delta \bm{z}^+_\lambda \times \Delta \bm{z}^-_\lambda |\rangle
\sim \overline{ \delta z}^2 \left(\frac{\lambda}{e^A L}\right)^{1 + (\beta - 1) \beta^{-\ln \beta}} \propto \lambda^{0.73},
\label{eq:ang_eq1} 
\end{equation} 
Using the same approach and setting 
\begin{equation}
\langle |\Delta \bm{z}^+_\lambda| |\Delta \bm{z}^-_\lambda| \rangle \sim \langle \delta z^+_\lambda \delta z^-_\lambda \rangle,
\label{eq:approx0} 
\end{equation} 
we obtain
\begin{equation}
\langle |\Delta \bm{z}^+_\lambda| |\Delta \bm{z}^-_\lambda| \rangle \sim \overline{ \delta z}^2
\left(\frac{\lambda}{e^A L}\right)^{1 + \ln \beta} \propto \lambda^{0.63},
\label{eq:ang_eq2} 
\end{equation} 
Combining Equations~(\ref{eq:ang_eq1}) and~(\ref{eq:ang_eq2}), we find that
\begin{equation}
\theta^\pm_\lambda \sim  \left(\frac{\lambda}{e^A L}\right)^{(\beta - 1)\beta^{-\ln\beta} - \ln\beta } \propto \lambda^{0.10}.
\label{eq:ang_eq3} 
\end{equation}

The above scalings reflect the contributions to $ \langle \delta
z^+_\lambda \delta z^-_\lambda \sin\theta_\lambda \rangle$ and $ \langle
\delta z^+_\lambda \delta z^-_\lambda \rangle$ from regions in which
$\delta z^+_\lambda \gg \delta z^\ast_\lambda$ or $\delta z^-_\lambda
\gg \delta z^\ast_\lambda$.  An upper bound on the contribution from
the remaining regions in which $\delta z^\pm \lesssim \delta
z^\ast_\lambda$ can be obtained by setting $\delta z^\pm_\lambda =
\delta z^\ast_\lambda$ and $\sin \theta_\lambda \sim 1$ in $
\langle \delta z^+_\lambda \delta z^-_\lambda \sin\theta_\lambda \rangle$
and $ \langle \delta z^+_\lambda \delta z^-_\lambda \rangle$.  The
resulting upper bounds become negligibly small compared to the values in
Equations~(\ref{eq:ang_eq1}) and (\ref{eq:ang_eq2}) as $\lambda /L
\rightarrow 0$, consistent with our assumption that the
large-$\delta z^\pm$ regions make the dominant contributions to $
\langle \delta z^+_\lambda \delta z^-_\lambda \sin\theta_\lambda \rangle$
and $ \langle \delta z^+_\lambda \delta z^-_\lambda \rangle$.

In the inertial range, the dominant contribution to $\langle (\delta
z^\pm_\lambda)^2\rangle$ comes from regions in which~$\delta
z^\pm_\lambda$ is unusually large.  In most of these regions, $\delta
z^\pm_\lambda \gg \delta z^\mp_\lambda$ and $|\Delta \bm{v}_\lambda|
\simeq |\Delta \bm{b}_\lambda| \simeq \delta z^\pm_\lambda/2$. Keeping
only the contribution to $\langle |\Delta \bm{v}_\lambda| |\Delta
\bm{b}_\lambda|\rangle$ from the regions that make the dominant
contributions to $ \langle (\delta z^+_\lambda)^2\rangle$ and $
\langle (\delta z^-_\lambda)^2\rangle$, we obtain the estimate
$\langle \Delta v_\lambda \Delta b_\lambda \rangle \simeq \langle
(\delta z^\pm_\lambda)^2\rangle/2$.  Since $ \Delta \bm{v}_\lambda
\times \Delta \bm{b}_\lambda = \Delta \bm{z}^-_\lambda \times \Delta
\bm{z}^+_\lambda/2$, Equations~(\ref{eq:sfs}) and (\ref{eq:ang_eq1})
imply that
\begin{equation}
\theta^{(vb)}_\lambda \sim \left(\frac{\lambda}{e^A L}\right)^{\beta^2 + (\beta-1)\beta^{-\ln\beta}} \propto \lambda^{0.21}.
\label{eq:ang_eq4} 
\end{equation} 

\begin{figure}[t]
\centerline{
\includegraphics[width=7cm]{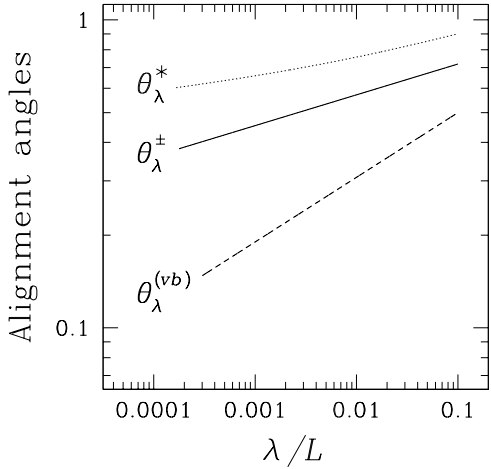}
}
\caption{Average alignment angles $\theta^\pm_\lambda$,
  $\theta^{(vb)}_\lambda$, and~$\theta^\ast_\lambda$ defined in
  Equations (\ref{eq:thetapm}), (\ref{eq:thetavb}),
  and~(\ref{eq:thetaast}). The
  angles $\theta^\pm_\lambda$ and $\theta^{(vb)}_\lambda$ scale as
  $\lambda^{0.10}$ and $ \lambda^{0.21}$, respectively. The angle
  $\theta^\ast_\lambda$ decreases as~$\lambda\rightarrow 0$ more
  slowly than any positive power of~$\lambda$.  For these plots, we
  set~$A=1$ in Equations~(\ref{eq:ang_eq3}) and (\ref{eq:ang_eq4}),
  where the constant~$A$ relates to the breadth of the PDF of~$\delta
  z^\pm_\lambda$ at the outer scale. We also set $A=1$ in
  Equation~(\ref{eq:chi1}) when evaluating $\mu_1$ in
  Equation~(\ref{eq:thetastar2}).
\label{fig:thetas} 
\vspace{0.2cm} 
}
\end{figure}

Finally, we define a third average alignment angle
\begin{equation}
\theta^\ast_\lambda = \left
\langle\frac{ |\Delta \bm{z}^+_\lambda \times \Delta \bm{z}^-_\lambda|}{
| \Delta \bm{z}^+_\lambda|| \Delta \bm{z}^-_\lambda|}\right\rangle.
\label{eq:thetaast} 
\end{equation} 
The angle~$\theta^\ast_\lambda$ is the volume average of the (sine of
the) angle between $\Delta \bm{z}^+_\lambda $ and $\Delta
\bm{z}^-_\lambda$, whereas $\theta^{\pm}_\lambda$ is a weighted
average of (the sine of) this angle that is dominated by regions in
which the fluctuation amplitudes are large.  If initially unaligned
$\delta z^+_\lambda$ and $\delta z^-_\lambda$ fluctuations collide and
$\delta z^+_\lambda \sim \delta z^-_\lambda$, then both fluctuations
evolve nonlinearly on the same timescale in an unpredictable and
disordered manner, which prevents the development of strong alignment.
Building upon this idea, we estimate $\theta^\ast_\lambda$ as
follows. We consider a new trial volume that is halfway between two
source regions, one for~$\delta z^+_\lambda$ and one for $\delta
z^-_\lambda$, which are separated by a distance~$2l^\ast_\lambda$.
Because this distance is twice as large as the typical parallel
correlation length that characterizes the bulk of the volume, we
take~$\delta z^\pm_\lambda$ in the two source regions to be
statistically independent. This assumption of statistical independence
breaks down for exceptionally strong fluctuations with large values
of~$l^\pm_\lambda$, but such fluctuations account for only a small
fraction of the volume and thus introduce only a small amount of error
into our estimate of~$\theta^\ast_\lambda$.  We then
set~$\theta_\lambda = 1$ in the trial volume if $\delta
z^+_{\lambda_1}$ and $\delta z^-_{\lambda_1}$ (where $\lambda_1 = e
\lambda$) in the two different source regions are equal to within a
factor of~3 (which is~$\simeq \beta^{-3}$), and otherwise we set
$\theta_\lambda = 0$.  We compare $\delta z^+_\lambda$ and $\delta
z^-_\lambda$ in the two source regions at scale~$\lambda_1$ because we
assume that the fluctuations cascade from scale~$\lambda_1$ to
scale~$\lambda$ as they propagate from the source regions to the trial
volume. This leads to the estimate
\begin{equation}
\theta^\ast_\lambda = e^{-2\mu_1}\sum_{q_1 = 0}^\infty \sum_{\substack{q_2 = 0 \\ |q_2 - q_1| \leq 3}}^\infty
\frac{\mu_1^{q_1 + q_2}}{q_1! \,q_2!},
\label{eq:thetastar1} 
\end{equation} 
where
\begin{equation}
\mu_1 = \mu - 1.
\label{eq:mu1} 
\end{equation} 
We rewrite Equation~(\ref{eq:thetastar1})  in the form
\begin{equation} 
\theta^\ast_\lambda = 
\sum_{q_1=0}^\infty \sum_{n=0}^{3} \frac{e^{-2\mu_1}\mu_1^{2q_1 + n}}{q_1! (q_1+n)!}
+  \sum_{q_2=0}^\infty \sum_{n=1}^{3} \frac{e^{-2\mu_1}\mu_1^{2q_2 + n}}{q_2! (q_2+n)!},
\label{eq:thetastar1.5} 
\end{equation} 
which is equivalent to
\begin{equation}
\theta^\ast_\lambda = e^{-2\mu_1}\sum_{n=-3}^{3} I_n(2\mu_1),
\label{eq:thetastar2} 
\end{equation} 
where $I_n(x)$ is the $n^{\rm th}$-order modified Bessel function of
the first kind.  As $\lambda\rightarrow 0$, 
\begin{equation}
\theta^\ast_\lambda \propto \left[
  \ln\left(\frac{L}{\lambda}\right)\right]^{-1/2},
\label{eq:thetastarlim} 
\end{equation} 
which decreases more slowly than any positive power of~$\lambda$.
We plot Equations~(\ref{eq:ang_eq3}),
(\ref{eq:ang_eq4}), and~(\ref{eq:thetastar2}) in
Figure~\ref{fig:thetas}.

\vspace{0.0cm} 
\subsection{Cross Correlation}
\label{sec:cross} 
\vspace{0.0cm} 

Equations~(\ref{eq:sfs}),
(\ref{eq:approx0}), and~(\ref{eq:ang_eq2}) yield the relation
\begin{equation}
\frac{\langle \delta z^+_\lambda \delta z^-_\lambda\rangle} {\langle
  \delta z^+_\lambda\rangle \langle \delta z^-_\lambda\rangle } \sim
\left(\frac{\lambda}{e^A L}\right)^{\ln\beta + 2\beta - 1} \propto
\lambda^{0.012},
\label{eq:dependence} 
\end{equation} 
which implies that $\delta z^+_\lambda$ and $\delta z^-_\lambda$
become anti-correlated at sufficiently small scales. However, this
anti-correlation grows extremely slowly as~$\lambda$ decreases.  This
very slow growth of anti-correlation results from the near
cancellation of two competing effects. First, we argued in
Equation~(\ref{eq:dzmxiast2}) that when~$\delta z^+_\lambda \gg \delta
z^\ast_\lambda$ in the trial volume in Figure~\ref{fig:source_region},
the likelihood that $\delta z^-_{\xi_\lambda} \gg \delta
z^\ast_{\xi_\lambda}$ in the source region is decreased because
intense, counter-propagating fluctuations rapidly annihilate.  This
effect acts to make $\delta z^+_\lambda$ and $\delta z^-_\lambda$
anti-correlated to an increasing degree as~$\lambda$ decreases,
because moments of the $\delta z^\pm_\lambda$ distribution are
increasingly dominated by exceptionally intense structures at smaller
scales (see, e.g., the discussion following
Equation~(\ref{eq:qnsigma})).  On the other hand, a large-amplitude,
coherent $\delta z^+_\lambda$ structure amplifies~$\delta z^-_\lambda$
to a value~$\sim \delta z^\ast_{\xi_\lambda}$ that exceeds~$\delta
z^\ast_\lambda$. Thus, a sheet-like coherent $\delta z^+_\lambda$
structure produces a weaker, sheet-like, $\delta
z^-_\lambda$-structure at the same location. On its own, this effect
would act to make $\delta z^+_\lambda$ and $\delta z^-_\lambda$
positively correlated, to a degree that would increase at smaller
scales, again because the moments of the $\delta z^\pm$ distribution
become increasingly dominated by large-amplitude fluctuations
as~$\lambda$ decreases.

\vspace{0.0cm} 
\section{Comparison with Numerical Simulations}
\label{sec:comp} 
\vspace{0.0cm}

The scalings in Section~\ref{sec:theory} agree reasonably well with a
number of results from direct numerical simulations.  For example, the
$k^{-1.52}$ scaling of the inertial-range power spectrum in our model
is in good agreement with the low-wavenumber ranges of the power
spectra in the numerical simulations of RMHD turbulence carried out
by~\cite{perez12} and \cite{beresnyak12,beresnyak14}. We note,
however, that \cite{beresnyak14} argued that the power spectra near
the dissipation scale vary with Reynolds number in his simulations in
the manner that would be expected if the inertial-range power spectrum
were proportional to~$k^{-5/3}$.  A detailed discussion of this point
is beyond the scope of this paper.

\begin{figure}[t]
\centerline{
\includegraphics[width=7cm]{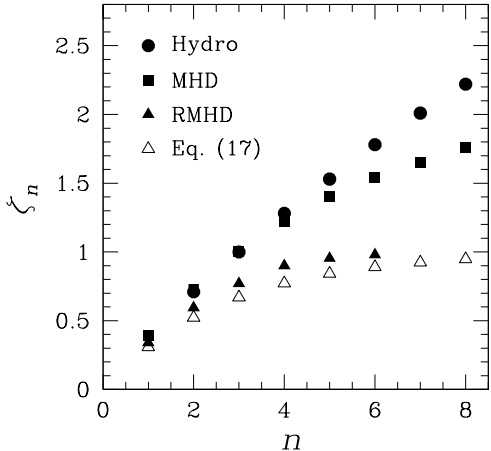}
}
\caption{Open triangles show the predicted scaling exponents~$\zeta_n$
  of the $n^{\rm th}$-order $z^\pm$ structure-function given in
  Equation~(\ref{eq:xin}).  Filled triangles show the scaling
  exponents for the velocity structure function in numerical
  simulations of RMHD turbulence~\citep{imazio13}.  Squares show the
  scaling exponents of the $z^\pm$ structure function in numerical
  simulations of 3D, non-helical, zero-mean-field, incompressible MHD
  turbulence~\citep{muller00}.  Circles show experimentally measured
  scaling exponents for the velocity structure function in
  hydrodynamic turbulence~\citep{benzi93}.
\vspace{0.2cm} 
\label{fig:SF} }
\end{figure}

In Figure~\ref{fig:SF}, we show the scaling exponents~$\zeta_n$ in
Equation~(\ref{eq:xin}) as well as the velocity-structure-function
scaling exponents found by \cite{imazio13} in simulations of strong
RMHD turbulence. Both sets of exponents asymptote to a value~$\simeq
1$ as~$n$ increases to large values. In contrast, as shown in the
figure, the scaling exponents for hydrodynamic turbulence and
non-helical, globally isotropic (zero mean field), 3D, incompressible
MHD turbulence reach significantly larger values at large~$n$ and do
not appear to asymptote towards a constant value as~$n\rightarrow
\infty$.  We note that in the incompressible MHD simulations of
\cite{muller03}, as $\delta z^\pm_L/B_0$ decreases to values $\ll 1$,
$\zeta_n$ becomes increasingly similar to the RMHD results shown in
Figure~\ref{fig:SF}. In contrast to hydrodynamic turbulence, $\zeta_3$
need not be~1 in~MHD turbulence or RMHD turbulence, because the
average of the right-hand side of Equation~(\ref{eq:eps1}) is not the
third moment of $\delta z^+$, but instead a correlation function
involving both $\delta z^+$ and $\delta
z^-$~\citep{politano98,boldyrev09}.

\cite{perez12} found that $\theta^{(vb)}_\lambda$ scaled like a power law that
was slightly flatter than~$\lambda^{1/4}$ in their highest-resolution,
highest-Reynolds-number RMHD turbulence simulation (their simulation
RB3b --- see their Figure~5), consistent with
Equation~(\ref{eq:ang_eq4}).  \cite{beresnyak12} found peak values
of~$\simeq 0.1$ for the scaling exponent $(d\ln
\theta^\pm_\lambda)/d\ln\lambda$ in numerical simulations of~RMHD turbulence,
consistent with Equation~(\ref{eq:ang_eq3}). In this same numerical
study, \cite{beresnyak12} found peak values of~$\simeq 0.2$ for the
scaling exponent $(d\ln \theta^{(vb)}_\lambda)/d\ln\lambda$, in agreement with
Equation~(\ref{eq:ang_eq4}).  \cite{beresnyak06} carried out
simulations of incompressible MHD turbulence and found
that~$\theta^\ast_\lambda$ decreased very slowly with decreasing~$\lambda$,
remaining close to unity throughout the inertial range in their
simulations, consistent with Equation~(\ref{eq:thetastarlim}).

\vspace{0.0cm} 
\section{Discussion}
\label{sec:discussion} 
\vspace{0.0cm} 

Intermittency has qualitatively different effects upon the energy
cascades rates in hydrodynamic turbulence and RMHD turbulence.  In
hydrodynamic turbulence, an intense vorticity structure interacts with
itself.  The concentration of fluctuation energy into a decreasing
fraction of the volume as $\lambda$ decreases thus reduces the energy
cascade timescale in the energetically dominant regions, to an
increasing degree as~$\lambda \rightarrow 0$.  Intermittency in
hydrodynamic turbulence thus acts to steepen the inertial-range power
spectrum. For example, $E(k) \propto k^{-1.71}$ in the She-Leveque
model, whereas $E(k) \propto k^{-5/3}$ in Kolmogorov's (1941)
theory. \nocite{k41,she94} In RMHD, since only counter-propagating
fluctuations interact, the concentration of~$\delta z^+_\lambda$
energy into a tiny fraction of the volume makes it difficult for a
$\delta z^-_\lambda$ fluctuation to ``find'' and interact with the
dominant~$\delta z^+_\lambda$ fluctuations. This in turn increases the
energy cascade timescale, to an increasing degree as
$\lambda\rightarrow 0$, causing the inertial-range power spectrum to
flatten relative to models of RMHD turbulence that neglect
intermittency, a point first made by~\cite{maron01}.

Like \cite{she94} and \cite{she95}, we assume that the fluctuation
amplitudes have a log-Poisson PDF and make an assumption about the
dimension of the most intense structures. On the other hand, the PDF
of the fluctuation amplitudes (and the PDF of the dissipation rate) in
the She-Leveque model has three parameters, whereas the PDF in our
model has just two: $\beta$ and~$\mu$ (Equations~(\ref{eq:z1}) and
(\ref{eq:Pq})).  (We do not count the overall normalization of the
fluctuation amplitudes --- e.g., $\overline{ \delta z}$ --- as a
parameter of the PDF in either model, because this normalization does
not affect the inertial-range scalings.) Our PDF has one fewer
parameter because of our argument that highly imbalanced collisions
reduce a fluctuation's length scale without affecting its amplitude,
which implies that the amplitude of the most intense ($q=0$)
fluctuations is independent of~$\lambda$. In order to determine the
extra free parameter in their model, \cite{she94} introduced an extra
assumption concerning the scaling of the energy dissipation rate of
the most intense structures.

In this paper, we draw heavily upon Boldyrev's (2005, 2006)
\nocite{boldyrev05,boldyrev06} argument that alignment within the
field-perpendicular plane plays an important role in the energy
cascade.  However, our treatment of scale-dependent dynamic alignment
differs from Boldyrev's.  In his theory, there is a single
characteristic alignment angle at each scale. In our model, at each
scale $\theta_\lambda$ varies systematically with the fluctuation
amplitudes (Equation~(\ref{eq:thetapm0})).  \cite{boldyrev06} argued
that a larger fluctuation amplitude reduces alignment. In our model,
given a scale~$\lambda$, larger fluctuation amplitudes are associated
with enhanced alignment, a phenomenon observed by \cite{beresnyak06}
in numerical simulations of incompressible MHD turbulence.  Also, in
our model, there are two distinct mechanisms for aligning $\Delta
\bm{v}_\lambda$ and $\Delta \bm{b}_\lambda$ fluctuations in regions
where the fluctuation amplitudes are large.  First, intense~$\delta
z^\pm_\lambda$ fluctuations rotate weaker~$\delta z^\mp_\lambda$
fluctuations into alignment, as illustrated in
Figure~\ref{fig:shear_slab}, which reduces~$\theta^{(vb)}_\lambda$
because $ \Delta \bm{v}_\lambda \times \Delta \bm{b}_\lambda = \Delta
\bm{z}^-_\lambda \times \Delta \bm{z}^+_\lambda/2$. Second, when the
fluctuations are intermittent, the turbulence becomes locally
imbalanced at small scales~\citep[cf.][]{perez09a}, with either
$\delta z^+_\lambda \gg \delta z^-_\lambda$ or $\delta z^-_\lambda \gg
\delta z^+_\lambda$ in the regions containing most of the fluctuation
energy. In such locally imbalanced regions, the velocity and
magnetic-field fluctuations are nearly parallel or anti-parallel,
regardless of whether $\bm{z}^+$ and $\bm{z}^-$ are
aligned~\citep{grappin13,wicks13a,wicks13b}. This second effect is why
$\theta^{(vb)}_\lambda$ decreases more quickly
than~$\theta^\pm_\lambda$ as~$\lambda/L$ decreases to small values.

\cite{grauer94}, \cite{politano95}, and \cite{muller00} developed
models of intermittent, incompressible, MHD turbulence based on the
approach of \cite{she94} and the assumption that $\epsilon^\pm_\lambda
\sim (\delta z^\pm_\lambda)^4/(\lambda v_{\rm A})$. \cite{muller00}
also developed a She-Leveque-like model of incompressible MHD
turbulence under the assumption that $\epsilon^\pm_\lambda \sim
(\delta z^\pm_\lambda)^3/\lambda$.  A major difference between our
approach and these previous studies is that we set
$\epsilon^\pm_\lambda \sim (\delta z^\pm_\lambda)^2 \delta
z^\mp_\lambda (\sin \theta_\lambda)/\lambda$, accounting for
alignment and treating $\delta z^+_\lambda$ and $\delta z^-_\lambda$
as separate but correlated random variables.

\vspace{0.0cm} 
\section{Conclusion}
\label{sec:conclusion} 
\vspace{0.0cm} 

We have constructed an analytic model of intermittent,
three-dimensional, strong RMHD turbulence that incorporates a new
phenomenology of scale-dependent dynamic alignment.  We restrict our
analysis to the case of ``globally balanced'' turbulence, in which the
cross helicity is zero.  There are three main assumptions in our
model. First, we take the fluctuation amplitudes to have a
scale-dependent, log-Poisson PDF.  In Section~\ref{sec:statistical},
we describe how this assumption can be motivated by treating a
fluctuation's evolution as a random, quantized, multiplicative
process, as in the work of \cite{she95}.  Second, we assume that the
most intense $\delta z^\pm_\lambda$ fluctuations are two-dimensional
current/vorticity sheets with a volume filling factor~$\propto
\lambda$. Third, we assume that the turbulence obeys a refined
similarity hypothesis (Equation~(\ref{eq:eps1})) that includes the
effect of dynamic alignment.

We argue that the largest contribution to the average $z^+$ cascade
power at any inertial-range scale~$\lambda$ comes from regions in
which $\delta z^+_\lambda \gg \delta z^-_\lambda$ and $\delta
z^+_\lambda \gg \delta z^\ast_\lambda$, where $\delta z^\ast_\lambda$
is the typical (median) fluctuation amplitude at scale~$\lambda$.  We
then develop an approximate theory describing how a large-amplitude,
coherent $\delta z^+_\lambda$ structure interacts with a much weaker
$z^-$ fluctuation. We show that during such an interaction, the $z^-$
fluctuation cascades rapidly to smaller scales without a reduction in
amplitude and rotates into alignment with the coherent $\delta
z^+_\lambda$~structure.  By accounting for these effects, we compute the
average $z^+$ cascade power using the assumed log-Poisson PDF
of~$\delta z^\pm_\lambda$.

This log-Poisson PDF has two free parameters, $\mu$ and~$\beta$ (see
Equations~(\ref{eq:z1}) and (\ref{eq:chi1})).  Our assumption that the
most intense fluctuations form two-dimensional structures with a
filling factor~$\propto \lambda$ determines~$\mu$ up to an additive
constant~$A$, which affects neither the power-law scalings in our
model nor the fact that~$\theta^\ast_\lambda$
(Equation~(\ref{eq:thetaast})) decreases logarithmically as~$\lambda
\rightarrow 0$.  The condition that the average cascade power is
independent of~$\lambda$ then determines~$\beta$.  Once we have
determined $\mu$ and~$\beta$, we compute the scalings of the $z^\pm$
power spectrum, higher-order structure functions, and three different
average alignment angles. Given the assumptions stated above, the
scalings in our model do not depend upon free parameters and agree
reasonably well with previously published numerical results.

There are a number of ways in which our model could be improved.  As
presented, our model can approximate a broad distribution of
outer-scale fluctuation amplitudes through the parameter~$A$, but the
outer-scale distribution is then forced to be log-Poisson.  A more
realistic approach might be to allow the quantity $\overline{ \delta
  z}$ (Equation~(\ref{eq:z1})) to be random with a distribution that
could be adjusted so as to model different forcing mechanisms in
forced turbulence or different initial conditions in decaying
turbulence.  Our finding that $\tau_{{\rm nl},\lambda}^\pm$ is an
increasing function of~$\delta z^\pm_\lambda$ at each scale suggests
that, at least for some dissipation mechanisms such as Laplacian
viscosity and resistivity, the dissipation scale is an increasing
function of fluctuation amplitude. This would mean that the unusually
intense fluctuations that make the dominant contribution to the power
spectrum begin dissipating at a larger scale than the fluctuations
that fill most of the volume. A useful direction for future research
would be to develop this idea further by exploring the consequences of
intermittency for the transition between the inertial and dissipation
ranges within the framework of our analytic model.  It would also be
useful to extend our model to allow for nonzero cross helicity in
order to investigate how intermittency affects strong ``imbalanced''
RMHD turbulence.  Finally, inhomogeneity of the background plasma can
fundamentally alter RMHD turbulence by causing the non-WKB reflection
of Alfv\'en waves~\citep{heinemann80}. This linear coupling between
counter-propagating Alfv\'en waves occurs in the solar atmosphere and
solar wind~\citep{dmitruk02,cranmer05,verdini07,chandran09c} and can
modify the power spectrum and energy-cascade timescales in solar-wind
turbulence~\citep{velli89,verdini12,perez13}. Extending our model to
account for background inhomogeneity and non-WKB wave reflection would
be helpful for understanding intermittent turbulence in the inner
heliosphere.

\acknowledgements This work was supported in part by grant NNX11AJ37G
from NASA's Heliophysics Theory Program, NASA grant NNN06AA01C to the
Solar Probe Plus FIELDS Experiment, NASA grant NNX13AF97G, and NSF
grant AGS-1258998. B.~Chandran was supported in part by a Visiting
Research Fellowship from Merton College, University of Oxford, and a
sabbatical leave from the University of New Hampshire.

\appendix
\vspace{0.0cm} 
\section{Highly Imbalanced Collisions}
\label{ap:shear} 

In this Appendix, we consider ``highly imbalanced collisions'' between
a large-amplitude, sheet-like, coherent, $\delta z^+_\lambda$
structure and smaller-amplitude $z^-$ fluctuations. We begin by
considering the effects of such collisions on the weaker, $z^-$
fluctuations. For this part of our analysis, we make the simplifying
approximation that the $\bm{z}^+$ field has the form of a linear shear
within the volume of the $\delta z^+_\lambda$ structure. We use the
term ``linear'' to refer to the functional form of~$\bm{z}^+$ in
Equation~(\ref{eq:zplusshear}) below, and not to imply that the
amplitude~$\delta z_\lambda^+$ is small.  We further assume that the
evolution of~$\bm{z}^-$ within the volume of the coherent~$\delta
z^+_\lambda$ structure does not depend strongly on the properties of
the $\bm{z}^+$ field outside of the structure.  This assumption allows
us to choose a convenient form for~$\bm{z}^+$ throughout all of space,
\begin{equation}
\bm{z}^+ = S(z, t) x \,\bm{\hat{y}},
\label{eq:zplusshear} 
\end{equation} 
where~$S(z,t)$ is the shearing rate and $(x,y,z)$ are Cartesian
coordinates chosen so that $\bm{B}_0$ is in the $z$~direction.
The RMHD equations can be rewritten in the form~\citep{schekochihin07}
\begin{equation}
\frac{\partial }{\partial t} \nabla_\perp^2 \psi^\pm \mp v_{\rm
  A}\frac{\partial }{\partial z}\nabla_\perp^2\psi^\pm =
-\frac{1}{2}\left(\{\psi^+,\nabla_\perp^2 \psi^-\} +
\{\psi^-,\nabla_\perp^2 \psi^+\} \mp
\nabla_\perp^2\{\psi^+,\psi^-\}\right),
\label{eq:RMHD2} 
\end{equation} 
where $\nabla_\perp = \bm{\hat{x}} \partial/\partial x + \bm{\hat{y}}
\partial/\partial y$, $\{g, h\} = \bm{\hat{z}} \cdot (\nabla_\perp g
\times \nabla_\perp h)$ for any functions $g$ and~$h$, and $\psi^\pm$
are the Els\"asser stream functions, which satisfy $\bm{z}^\pm =
\bm{\hat{z}}\times \nabla_\perp \psi^\pm$.
Equation~(\ref{eq:zplusshear}) then implies that $\psi^+ = Sx^2/2$ to
within an arbitrary additive function of the~$z$ coordinate and~time.
Upon substituting this value of~$\psi^+$ into
Equation~(\ref{eq:RMHD2}), we obtain
\begin{equation}
\left(\frac{\partial }{\partial t} + Sx \frac{\partial
}{\partial y} + v_{\rm A} \frac{\partial}{\partial z}\right)
\nabla_\perp^2 \psi^- = - S\,
\frac{\partial^2\psi^-}{\partial x\partial y}.
\label{eq:RMHD3} 
\end{equation} 
Although $\psi^+ = Sx^2/2$ is not localized, 
we take~$\psi^-$ to vanish sufficiently rapidly as~$x^2 + y^2 \rightarrow \infty$ that 
\begin{equation}
f = \frac{1}{(2\pi)^2} \int dx dy \psi^- e^{-ik_x x - i k_y y}
\label{eq:deff} 
\end{equation} 
is defined and the Fourier transforms in $x$ and~$y$ of each term in
Equation~(\ref{eq:RMHD3}) are defined.  The Fourier transform of
Equation~(\ref{eq:RMHD3}) yields
\begin{equation}
 \left(\frac{\partial}{\partial t} - Sk_y \frac{\partial}{\partial k_x}
+ v_{\rm A} \frac{\partial }{\partial z}\right) (k_\perp^2 f) = 
- S k_x k_y f.
\label{eq:RMHDft} 
\end{equation} 
To solve Equation~(\ref{eq:RMHDft}), we define a family of
trajectories in $k_x-z$ space through the equations $dk_x/dt = -
Sk_y$ and $dz/dt = v_{\rm A}$. The total time derivative of any
function $G(k_x(t), k_y, z(t),t)$ along one of these trajectories is then
$(d/dt)G = ( \partial/\partial t - Sk_y \partial/\partial k_x + v_{\rm
  A} \partial/\partial z)G$.  Since $(d/dt) k_\perp^2 = -2Sk_x k_y$,
we can rewrite Equation~(\ref{eq:RMHDft}) as $(d/dt)(k_\perp f) =
0$. The solution to Equation~(\ref{eq:RMHDft}) is thus
\begin{equation}
f(k_x, k_y, z, t) = \frac{k_{\perp 0}}{k_\perp}\, f_0(k_{x0}, k_y, z_0),
\label{eq:solution1} 
\end{equation} 
where $z_0 = z - v_{\rm A} t$, $k_{x0} = k_x + k_y H$,
\begin{equation}
H = \int_0^{t} S(z_0 + v_{\rm A} t^\prime, t^\prime)dt^\prime,
\label{eq:defH}
\end{equation}
$f_0(k_x, k_y, z) = f(k_x, k_y, z, 0)$, $k_\perp = \sqrt{k_x^2 + k_y^2}$, and
$k_{\perp 0} = \sqrt{k_{x0}^2 + k_y^2}$.  The Fourier transform
of~$\bm{z}^-$ is then
\begin{equation}
\bm{z}^-_k = i (\bm{\hat{z}}\times \bm{\hat{k}_\perp}) k_{\perp 0} f_0(k_{x0}, k_y, z_0),
\label{eq:zsolve} 
\end{equation} 
where $\bm{\hat{k}_\perp} = (k_x \bm{\hat{x}} +
k_{y}\bm{\hat{y}})/k_\perp$.

If we focus on a cross section of~$\psi^-$ in the $xy$-plane, then
this cross section is advected to larger~$z$ at speed~$v_{\rm A}$, and
each Fourier component of~$\psi^-$ in that cross section is advected
in~$k_x$ at ``wavenumber velocity'' $-Sk_y$, resulting in the equation
\begin{equation}
k_x = k_{x0} -  k_y H,
\label{eq:kxt} 
\end{equation} 
where~$k_{x0}$ is the initial value of~$k_{x}$.
The amplitude $k_\perp f$
of each Fourier component of~$\bm{z}^-$ in this (moving) cross section
remains constant in
time. Moreover, as $t$ increases, $|k_x|$ may initially decrease, but
eventually $|k_x|$ increases without bound.  This
causes~$\bm{\hat{k}_\perp}$ to align with the~$x$ axis and~$\bm{z}^-$
to align with the~$y$ axis. The angle~$\theta$ between~$\bm{z}^-$
and~$\bm{z}^+$ is $\sin^{-1}|k_{y}/k_\perp|$.  If $k_{x0} \sim k_{y}$
and $H \gg 1$, then $k_x \simeq -k_{y}H$,
\begin{equation}
\theta \simeq H^{-1} \hspace{0.6cm}  \mbox{ and }
\hspace{0.6cm} 
\frac{k_\perp}{k_{\perp 0}} \simeq  H.
\label{eq:thetakp} 
\end{equation} 

We now use these results to obtain an approximate description of the
evolution of~$z^-$ fluctuations as they propagate a
distance~$l^\ast_\lambda$ from the source region to the trial volume
depicted in Figure~\ref{fig:source_region} in the limit that $\delta
z^+_\lambda \gg \delta z^\ast_\lambda$.  Because the $z^-$
fluctuations in the source region have not yet interacted with the
coherent~$\delta z^+_\lambda$ structure depicted in
Figure~\ref{fig:source_region}, they do not yet ``know about'' the
orientation of this structure. The typical case is thus that $k_{x0}
\sim k_{y}$, so that the $z^-$ fluctuations are not initially aligned
with the $\delta z^+_\lambda$ structure.  We set $S \rightarrow \delta
z^+_\lambda/\lambda$, which is the shearing rate associated with the
$\delta z^+_\lambda$ structure, and we set $t\rightarrow l^\ast_\lambda/v_{\rm
  A}$, which is the time it takes the $z^-$ fluctuations to propagate
from the source region to the trial volume.  This leads to $ H =
\delta z^+_\lambda/\delta z^\ast_\lambda \gg 1$, so that
Equation~(\ref{eq:thetakp}) applies.  The condition $k_\perp f =
\mbox{ constant}$ and Equation~(\ref{eq:thetakp}) then lead to
Equations~(\ref{eq:thetapm0}) and (\ref{eq:tvsr1}).
Equation~(\ref{eq:thetakp}) also implies that the perpendicular length
scale of the~$\bm{z}^-$ fluctuations decreases by a factor of~$\delta
z^+_\lambda/\delta z^\ast_\lambda$ as the $z^-$ fluctuations propagate
from the source region to the trial volume.

Finally, we consider how highly imbalanced collisions affect the
larger-amplitude, coherent~$\delta z^+_\lambda$ structure. As can be
seen in the bottom half of Figure~\ref{fig:shear_slab}, the $\delta
z^-_\lambda$ fluctuations within a sheet-like coherent~$\delta
z^+_\lambda$ structure have been sheared in such a way that they
resemble a smaller-amplitude, counter-propagating, current/vorticity
sheet that is nearly aligned with the coherent~$\delta z^+_\lambda$
structure. The nature of the effect of this $\delta z^-_\lambda$
current/vorticity sheet on the original $\delta z^+_\lambda$ structure
is also effectively linear shearing. Because of this, we can repeat
the arguments leading from Equation~(\ref{eq:zplusshear}) to
Equation~(\ref{eq:kxt}), interchanging the roles of~$z^+$
and~$z^-$. We thus conclude that highly imbalanced collisions between
a sheet-like coherent $\delta z^+_\lambda$ structure and much weaker
$z^-$ fluctuations change the scale but not the amplitude of
the~$\delta z^+_\lambda$ structure, as argued in
Section~\ref{sec:statistical}. We also note that the scale of the
$\delta z^+_\lambda$ structure can either increase or decrease,
depending on the directions of the vector fluctuations in the
``colliding'' fluctuations (i.e., depending on the relative signs of
the two terms on the right-hand side of~Equation~(\ref{eq:kxt}), when
we have interchanged the roles of~$z^+$ and~$z^-$ in
Equations~(\ref{eq:zplusshear}) through (\ref{eq:kxt}) in order to
describe the evolution of~$\delta z^+_\lambda$).

\bibliography{articles}

\end{document}